\documentstyle[twocolumn,epsf,aps]{revtex}

\draft

\begin{document}

\title{Nonequilibrium Steady State in a Quantum System:\\
One-dimensional Transverse Ising Model with Energy Current 
\vspace{0.4truecm}
}

\author{Tibor Antal and Zolt\'an R\'acz}

\vskip 0.5truecm

\address{
Institute for Theoretical Physics, E\"otv\"os University\\
1088 Budapest, Puskin u. 5-7, Hungary
}
\author{L\'aszl\'o Sasv\'ari}

\vskip 0.5truecm

\address{
Department of Solid State Physics, E\"otv\"os University\\
1088 Budapest, M\'uzeum krt. 6-8, Hungary
\vspace {0.5truecm}
}

\vspace {1truecm}

\address{
\centering{
\medskip\em
\begin{minipage}{14cm}
{}We study the nonequilibrium steady states of an Ising chain 
in a transverse field, $h$, by investigating the effect of
a field, $\lambda$, which drives the current of energy. 
The zero-temperature, $h-\lambda$ phase diagram is determined
exactly and it is found that the energy 
current appears continuously above a 
threshold, $\lambda >\lambda_c(h)$. The long-range magnetic order 
(present for $h<h_c, \lambda <\lambda_c$) is destroyed in the 
current-carrying state
where the correlations are characterized by
power-law, oscillatory decay. The mechanism which generates the power-law 
correlations in the current-carrying state is discussed.
\pacs{\noindent PACS numbers: 05.50.+q, 05.70.Ln, 75.10.Jm}
\end{minipage}
}
}

\maketitle
\narrowtext

Nonequilibrium steady states have been much investigated but 
a general theory comparable to the theory of equilibrium systems is 
still lacking.
A traditional approach to nonequilibrum phenomena 
starts with the construction of 
deterministic phenomenological equations, followed by a linear or nonlinear 
stability analysis of the solutions and ends up with the study of the effect 
of fluctuations by adding noise terms to the equations \cite{{Cross},{Zia}}. 
An alternative and somewhat more microscopic approach
is the introduction of kinetic Ising models where 
the nonequilibrium steady states are produced by
the spins being driven by external fields 
or by being in contact with several heat baths at different temperatures 
\cite{{KatzLebo},{Marro}}. 
 
Both of these approaches have lead to successful descriptions of a number of 
particular phenomena but they have not been very helpful in 
deducing general conclusions. The main problem is that the observed 
properties of nonequilibrium steady states 
are strongly dependent on the details of the
dynamics (even the presence or absence of a phase transition or the type of
the phase transitions, etc. may depend
on the details of the transition probabilities). From a theoretical 
point of view, the problem is further complicated 
by the absence of
the detailed balance condition which provides at least 
some constraints on the possible
forms of dynamics of near-equilibrium fluctuations. 

  A way to avoid the above problem of too much freedom is to investigate
systems which, in contrast to models such as the Ising model, have natural
(quantum mechanical) dynamics.  A nontrivial model with
intrinsic dynamics investigated in this paper is 
the one-dimensional ($d=1$) transverse
Ising model. Once a quantum mechanical
model is chosen, the next question is how to make it a nonequilibrium
system.  A possible route is to couple it to heat baths
just as in the case of the various kinetic Ising models.  
Unfortunately, the coupling of quantum mechanical
systems to classical heat baths is not quite understood and
leads again to much arbitrariness.  In order to avoid this, we
follow a different path by noting that nonequilibrium steady
states are always associated with some kind of current
(of heat, particle, momentum, etc.).  Thus we make a
nonequilibrium steady state in a quantum system by imposing
a current on the system.  For example, in case of the
transverse Ising model where the only conserved quantity is
the energy, we shall constrain the system to have an energy current. 
It is hard to treat, however, a `microcanonical' type constraint of
fixed energy flow and so,
we shall actually use the easier, `canonical' description which consists of 
introducing a field which drives the given (energy) current. 

The next
question is what to investigate.  An obvious choice is to
study the changes produced by the presence of a current in 
the orderings and in the phase
diagram of the system.  In particular, for the $d=1$
transverse Ising model, one can investigate whether the $T=0$ Ising
transition 
remained in the same universality class and whether new phases 
emerged as a consequence of the energy current.
Furthermore, it is known that currents in nonequilibrium steady states 
usually generate long-range correlations \cite{Zia2}.
Thus it is natural to inquire if similar phenomena would occur in a quantum 
nonequilibrium steady state. 

   The practical realization of the above program consists of the 
following steps: 

  i) The starting point is the Hamiltonian which, for the
$d=1$ transverse Ising model, has the form:  
\begin{equation}
\hat H_I=-\sum_\ell \sigma^x_\ell\sigma^x_{\ell+1} -\frac{h}{2} \sum_\ell 
\sigma^z_\ell
\label{H_I}
\end{equation} 
where the spins are represented by Pauli spin matrices
$\sigma^\alpha_\ell$ ($\alpha =x,y,z$) at sites
$\ell=1,2,...,N$ of a $d=1$ periodic chain
($\sigma^\alpha_{N+1}=\sigma^\alpha_1$), and $h$ is the
transverse field in units of the Ising coupling.

  ii) The local energy current, $\hat J_\ell$, is calculated by taking a
time derivative of the energy density, using the quantum 
mechanical equation of motion, and representing the 
result as a divergence of the energy current ($\hbar =1$ is used): 
\begin{equation}
\hat J_\ell = \frac{h}{4}\sigma^y_\ell(\sigma^x_{\ell-1}-\sigma^x_{\ell+1})
\quad .
\label{current}
\end{equation} 

  iii) The `macroscopic' 
current $\hat J=\sum_\ell \hat J_\ell$ is added to $\hat H_I$
with a Lagrange multiplier, $-\lambda$, 
\begin{equation}
\hat H =\hat H_I -\lambda \hat J \quad .
\label{ham}
\end{equation}
Note that the energy current, $\hat J$, is associated with $\hat H_I$
and not with the new Hamiltonian, $\hat H$.
We also emphasize that $\hat H$ is just another equilibrium Hamiltonian, 
it differs from $\hat H_I$ by an extra term which breaks the left-right 
symmetry of $\hat H_I$. 
Finding the ground state of $\hat H$, however, gives us the minimum energy 
state of $\hat H_I$
which carries an energy current, $J =\langle \hat J/N \rangle$  
(brackets $\langle\rangle$ denote the expectation value in the ground-state 
of $\hat H$). Thus the ground-state
properties of $\hat H$ provide us with the properties of the nonequilibrium 
steady states of the transverse Ising model.

  iv) The correlations and ordering properties of the system are investigated
as a function of $\lambda$. Of course, 
one can also find that value of $\lambda$ which generates a given 
energy current, $J =\langle \hat J/N \rangle$, in the transverse Ising model
and then all steady state properties can be obtained in terms of $J$.

It should be noted that the terms in $\hat J$ can be 
rearranged so that the 
energy current becomes a current through the bonds:
\begin{equation}
\hat J =\sum_\ell \hat{\bar J_\ell} = \frac{h}{4}\sum_\ell 
(\sigma^x_\ell\sigma^y_{\ell+1}-\sigma^y_{\ell}\sigma^x_{\ell+1})\quad .
\label{bondcurr}
\end{equation} 
This form is remarkable because $\hat{\bar J_\ell}$ can be recognized 
as the Dzyaloshinskii-Moriya interaction in the theory of weak ferromagnetism 
\cite{DM}. Moreover, the current of the 
$z$ component of the magnetization in the 
transverse XY model is also of the above form \cite{ARSch}, and similar terms 
(apart from an overall factor of $i$ which 
makes the Hamiltonian nonhermitian) appear in the description of 
driven diffusive systems \cite{Stinchcombe},
as well as in the Hamiltonian description of the direct electric field
of the 6-vertex model \cite{denNijs}.  

\begin{figure}[htb]
\centerline{
	\epsfxsize=9.0cm
	\epsfbox{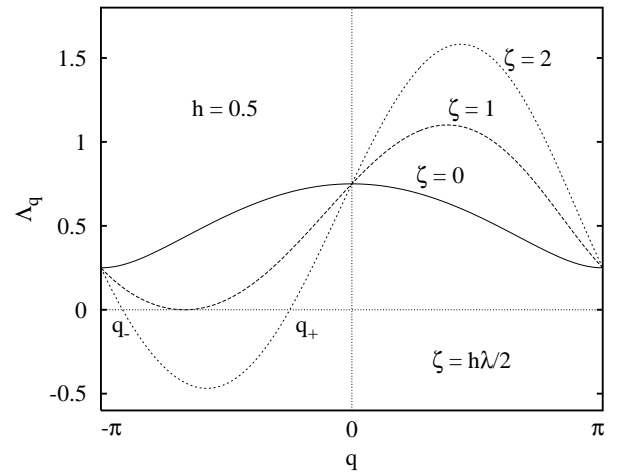}
		\vspace{0.5cm}
		   }
\caption{Spectrum of the transverse Ising model 
in the presence of a field ($\lambda$) which drives the current of energy. 
The excitation energies are given as $\omega_q=|\Lambda_q|$. 
The qualitative picture is the same at all transverse fields 
$h$ except at $h=1$
where the $q=\pm\pi$ gap disappears. 
}
\label{fig:spectrum}
\end{figure}

We turn now to the solution of the problem. 
Since $[\hat H_I, \hat J]=0$, the Hamiltonian $\hat H$ is diagonalized 
using the same transformations which diagonalize 
$\hat H_I$ \cite{{LSM},{Pfeuty}}: first,
creation-annihilation operators are introduced ($a_\ell^{\pm}=\sigma_\ell^x\pm 
i\sigma_\ell^y$), 
then the Jordan-Wigner transformation \cite{LSM} is used to transform them 
into fermion operators 
($c_\ell, c_\ell^+$) and,
finally, a Bogoljubov transformation \cite{Bogo} is employed 
on the $q$ and $-q$ components of the 
Fourier transforms of $c_\ell$-s. As a result one finds the spectrum of 
excitation energies as $\omega_q=\left|\Lambda_q\right|$ where 
\begin{equation}
\Lambda_q=\frac{1}{2}\left( \sqrt{1+h^2+2h\cos q} +\zeta \sin q 
\right) \quad .
\label{excitations}
\end{equation} 
Here the wave numbers are restricted to 
$-\pi\le q\le\pi$ in the thermodynamic limit ($N\rightarrow \infty$) and 
we introduced $\zeta = \lambda h/2$ which appears to be the natural variable 
instead of $\lambda$. Fig.\ref{fig:spectrum} displays the spectrum 
for $h=0.5$ and various $\zeta$ and one can see that
the $q\rightarrow -q$ symmetry of the spectrum is broken for $\zeta \not= 0$. 
For small $\zeta$ when $\Lambda_q\ge 0$, however,  
the ground-state remains that of the
transverse Ising model ($\zeta=0$) and, accordingly, no energy current 
flows $(J=0)$. This rigidity of the ground state 
against the symmetry-breaking field which drives the energy current is a
consequence of the facts that the fermionic spectrum of the 
transverse Ising model has a gap and 
that the operator $\hat J$ commutes with $\hat H_I$.
Actually, the rigidity persists even at the critical point $h_c=1$ where 
the gap disappears. The reason for this is more subtle and 
related to the fact that the critical spectrum of $\hat H_I$ 
has a finite slope at $q=\pm \pi$.

\vspace{0.3cm}
\vspace{0.5cm}

The ground-state properties do change when $\Lambda_q<0$ in an interval 
$[q_-,q_+]$ since these $q$ states 
become occupied and $q_\pm$ 
depends on both $h$ and $\zeta$. The line $\zeta_c(h)$
which borders the region of unchanged transverse Ising behavior is 
obtained from the conditions
\begin{equation}
\Lambda_q =0 \quad , \quad \partial \Lambda_q/\partial q =0 \quad ,
\label{TIborder}
\end{equation} 
and the solution
\begin{equation}
\zeta_c=\frac{\lambda_c h}{2}= \cases { \zeta_c^+=h & \qquad $h\ge 1$ , \cr
                  \zeta_c^-=1 & \qquad $h< 1$ \cr} 
\label{TIborder2}
\end{equation} 
is displayed on the phase diagram (Fig.\ref{fig:phasedia}) as a solid line. 
Fig.\ref{fig:phasedia} also shows the phase boundary (dashed line) between the 
magnetically ordered ($h<1$, $\zeta<1$) and disordered ($h\ge 1$,
$\zeta<h$) transverse Ising regions. 
Since the ground state is independent of $\zeta$ for $\zeta<\zeta_c$, 
the transition across the dashed line is a second order transition 
belonging to the $d=2$ Ising universality class \cite{Pfeuty}.

In the following we shall argue that the
$\zeta>\zeta_c$ region can be considered as a distinct phase since i) the 
energy current is nonzero, ii) there is no long-range magnetic order, and iii)
the magnetic correlations are oscillatory with
amplitudes decaying as a power of distance.

\setlength{\unitlength}{1.0cm}
\begin{figure}

\null\hspace{0.5truecm}
\begin{picture}(7,6)
\thicklines
\put(0,0){\vector(1,0){7.0}}
\put(0,0){\vector(0,1){6.0}}
\put(3,0){\line(0,1){3}}
\multiput(0,3)(0.46,0){7}{\line(1,0){0.25}}
\put(3,3){\line(1,1){2.8}}

\put(5.6,-0.5){$\zeta= h \lambda / 2$}
\put(2.9,-0.5){$1$}
\put(-0.1,-0.5){$0$}
\put(-0.5,5.5){$h$}
\put(-0.4,2.9){$1$}
\put(-0.4,0){$0$}

\put(0.7,4.8){$J=0$}
\put(0.7,4.3){disordered}
\put(0.7,3.8){$\langle \sigma_\ell^x \rangle = 0 $}
\put(0.7,2){$J=0$}
\put(0.7,1.5){ordered}
\put(0.7,1){$\langle \sigma_\ell^x \rangle \ne 0 $}
\put(4.5,2.8){$J \ne 0$}
\put(3.7,2){$\langle \sigma_\ell^x \sigma_{\ell+n}^x \rangle \sim 
	{1 \over \sqrt n} \cos kn$}

\put(3.5,0.5){\vector(-1,0){0.5}}
\put(3.7,0.4){$k=0$}
\put(5,4.5){\vector(-1,0){0.5}}
\put(5.2,4.4){$k \ne 0$}
\end{picture}

\vspace{1truecm}
\caption{Phase diagram in the $h-\zeta$ plane where $h$ is 
the transverse field while $\zeta=h\lambda/2$ is 
the effective field which drives the current of 
energy.
Power-law correlations are present in the current-carrying phase ($J\not= 0$)
and on the Ising critical line (dashed line).
 }
\label{fig:phasedia}
\end{figure}
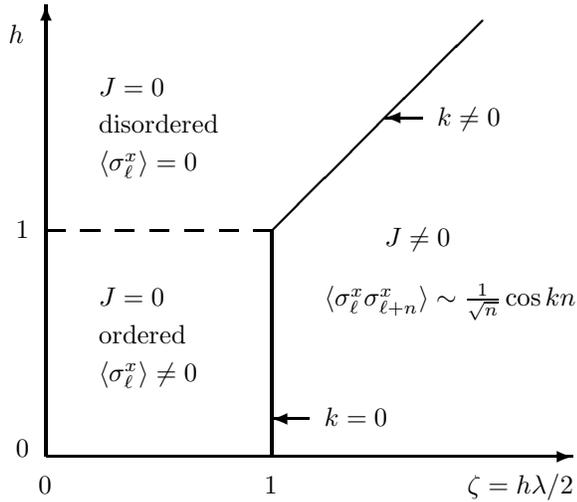
\vspace{0.5cm}

The energy current is
expressed as a product of two fermion operators and can be 
calculated exactly. It is zero for $\zeta \le\zeta_c$ while, 
for $\zeta >\zeta_c$, one obtains:  
\begin{equation}
J =\langle \hat J/N \rangle=\frac{h}{4\pi\zeta^2}\sqrt{(\zeta^2-h^2)
(\zeta^2-1)} \quad .
\label{J}
\end{equation}
At the phase boundary, $\zeta_c$, we have $J\rightarrow 0$ as 
\begin{equation}
J \sim  \cases { (\zeta-\zeta_c)^{1/2} & \qquad $h\not= 1$ , \cr
                     ~\zeta-\zeta_c & \qquad $h= 1$ .\cr} 
\label{J->0}
\end{equation}
Thus the $\zeta_c$ line can be considered as a line of second order transition 
provided $J$ is 
viewed as an order parameter. 

There are several other quantities which 
change nonanalytically as we cross into the current carrying phase. For
example, the change in the $z$ component of the magnetization at fixed 
$h$ is given by
\begin{equation}
\langle \sigma^z_\ell \rangle_\zeta -\langle \sigma^z_\ell 
                                \rangle_{\zeta_c}\sim  
             \cases { (\zeta-\zeta_c)^{1/2} & \qquad $h>1$ , \cr
                      ~\zeta-\zeta_c & \qquad $h=1$ ,\cr 
                      (\zeta-\zeta_c)^{3/2} & \qquad $h<1$ .\cr} 
\label{corr_z}
\end{equation}
The different exponents obtained for $h>1$ and $h<1$ reflect the fact that 
the $J\not=0$ phase neighbors magnetically ordered ($\langle \sigma^x_\ell
\rangle \not= 0$) and magnetically disordered ($\langle \sigma^x_\ell
\rangle = 0$) phases along the $\zeta_c^+$ and $\zeta_c^-$ portions 
of the $\zeta_c$ phase boundary.  

The long-range magnetic order 
($\langle \sigma^x_\ell\rangle \not= 0$) disappears for $\zeta>\zeta_c$.
This can be seen by investigating the 
$\langle \sigma^x_\ell \sigma^x_{\ell+n}\rangle$ 
correlations which can be expressed through Pfaffians \cite{{Fubini},{Baruch}}. 
In the presence of long-range order, one should have
$\langle \sigma^x_\ell \sigma^x_{\ell+n}\rangle\rightarrow 
\langle \sigma^x_\ell \rangle^2 \not=0$ for $n\rightarrow \infty$. 
Instead, we find that the correlations decay to zero at large distances as
\begin{equation}
\langle \sigma^x_\ell \sigma^x_{\ell+n}\rangle 
\sim\frac{Q(h,\zeta)}{\sqrt n}\cos(kn) 
\label{xxcorr}
\end{equation}
where the wavenumber, $k$, depends only on $\zeta$, 
\begin{equation}
k=\arccos{\zeta}^{-1} \quad .
\label{wavenumber}
\end{equation} 
The above results (\ref{xxcorr},\ref{wavenumber}) are exact 
in the $\zeta\rightarrow\infty$ limit where 
$Q(h,\infty)=e^{1/2}2^{-4/3}A^{-6}\approx 0.147$ is the amplitude of the 
$\langle \sigma^x_\ell \sigma^x_{\ell+n}\rangle$ correlations of the $d=1$ 
$XY$ model \cite{Baruch} ($A\approx 1.282$ is Glaisher's constant).
The connection to the 
$XY$ model and the  
details of the calculation will be published separately \cite{ARS}.

\vspace{0.3cm}
\begin{figure}
\centerline{
	\epsfxsize=9.0cm
	\epsfbox{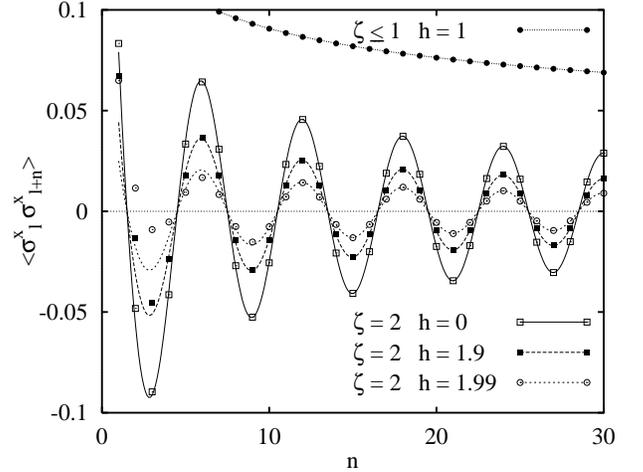}
		\vspace*{0.5cm}
		   }
\caption{The correlation function 
$\langle \sigma^x_\ell \sigma^x_{\ell+n}\rangle$ at various 
transverse fields $h$ with $\zeta=h\lambda/2$ fixed at $\zeta=2$.
The lines are fits to the large $n$ behavior which is assumed to be of the 
form $Qn^{-1/2}\cos kn$. Changes in the amplitude $Q$ occur 
mainly near the $\zeta_c^+$ phase boundary $(h\rightarrow 2)$ 
where $Q\rightarrow 0$. 
Note that the convergence to the asymptotic behavior becomes slower as 
$h\rightarrow 2$. For comparison, we have also plotted the correlations 
$\langle \sigma^x_\ell \sigma^x_{\ell+n}\rangle\sim n^{-1/4}$ on the Ising 
critical line (dashed line on Fig.2).
	}
\label{fig:xxcorr}
\end{figure}
\vspace{0.5cm}

For $\zeta_c<\zeta<\infty$, the expressions (\ref{xxcorr},\ref{wavenumber}) are
obtained from the numerical analysis of 
$\langle \sigma^x_\ell \sigma^x_{\ell+n}\rangle$ for 
finite $n\le 100$. Typical results are shown on Fig.\ref{fig:xxcorr}. 
The numerical results are behaved well enough so that, in addition to the 
$n$ dependence of $\langle \sigma^x_\ell \sigma^x_{\ell+n}\rangle$, 
the functional
form of the amplitude $Q(h,\zeta)$ can also be investigated. 
We find that the following 
expression 
fits the results everywhere ($\zeta >\zeta_c$) except close to the 
phase boundaries, $\zeta_c^\pm$, or near the $h=1$ line:
\begin{equation}
Q(h,\zeta)\approx Q(h,\infty) 
\left(\frac{\zeta^2-h^2}{\zeta^2-1}\right)^{1/4} \quad .
\label{amplitude1}
\end{equation}
Near either $\zeta=\zeta_c^\pm$ or $h=1$, 
the convergence to the asymptotic form (\ref{xxcorr}) is slow and
the numerics becomes unreliable. The convergence becomes fast 
again on the $h=1$ line itself where the following form appears to be valid:
\begin{equation}
Q(1,\zeta)\approx Q(h,\infty) 
\left(\frac{\zeta^2}{\zeta^2-1}\right)^{1/8}  \quad .
\label{amplitude2}
\end{equation} 

Power-law decay of correlations for $\zeta>\zeta_c$ is 
present in other physical 
quantities as well. For example, the envelopes of both 
$\langle \sigma^z_\ell\sigma^z_{\ell+n}\rangle$ and 
$\langle \hat J_\ell \hat J_{\ell+n}\rangle$ 
decay in the large $n$ limit as $n^{-2}$ \cite{ARS}. 
Thus we find power-law correlations in the current carrying state 
in agreement with the notion that power-law correlations are a 
ubiquitous feature of nonequilibrium steady states \cite{eqcorr}.

Power-law correlations in quantum models are usually associated with
a gapless excitation spectrum. Thus, provided the 
transverse Ising model can be considered as an instructive example, we 
can see a general connection between the emergence of power-law
correlations and the presence of a current. Indeed, let us
assume that a system with Hamiltonian $\hat H_0$ has a spectrum with a gap 
between the ground-state and the lowest excited state. Furthermore, let
$\hat J$ be a `macroscopic' current of a conserved quantity 
such that $[\hat H_0,\hat J]=0$. Generally, there is no current in the 
the ground state and adding $-\lambda\hat J$ to $\hat H_0$ 
does not change the $\langle \hat J \rangle =0$ result for small $\lambda$. 
Current can flow only if some excited states mix with the ground state  
and, consequently, 
a branch of the excitation spectrum must come down and intersect the 
ground-state energy in order to have $\langle \hat J \rangle \not= 0$. Once 
this happens, however, the gap disappears and one can expect power-law 
correlations in the current-carrying state. 

Admittedly, the above argument is not strict and is just a reformulation 
(in general terms) of the transverse Ising results. 
We believe, however, that the above picture is robust and suggestive enough to 
warrant further studies in this direction. 
  
In summary, we have seen that 
nonequilibrium steady states in quantum systems 
can be constructed by switching on fields which drive the 
current of conserved quantities. Studies of these states may 
be of importance since their construction is free from the arbitraryness 
inherent in classical nonequilibrium dynamics. On the 
particular example of the transverse Ising model, we could observe that there 
are power-law correlations in the current-carrying states and we argued that 
this may be a general feature of quantum systems.

\section*{Acknowledgments}

We thank Z. Horv\'ath, L. Palla, 
P. Ruj\'an, G.M. Sch\"utz, R. Stinchcombe, G. Szab\'o, and T. T\'el for helpful 
discussions. This work was supported by the 
Hungarian Academy of Sciences Grant OTKA T 019451, 
and by EC Network Grant ERB CHRX-CT92-0063.


\begin{references}
\bibitem{eqcorr}Just as in equilibrium systems, the correlations 
                decay exponentially in the $\zeta <\zeta_c$ states (except
                on the critical line, $h=h_c$). This is not surprising since 
                no steady-state current (local or global) flows in those
                states and thus they should be considered as equilibrium 
                states.
\end{references}
\end{document}